# Secure and Dependable Virtual Network Embedding

Luís Ferrolho (student), Max Alaluna (student), Nuno Neves, Fernando M. V. Ramos
{lferrolho,malaluna}@lasige.di.fc.ul.pt, nuno@di.fc.ul.pt, fvramos@ciencias.ulisboa.pt
*LaSIGE, Faculdade de Ciências, Universidade de Lisboa, Portugal*

Network virtualization has emerged as a powerful technique to allow multiple heterogeneous networks specified by different users to run on a shared infrastructure. A major challenge is how to make efficient use of the shared resources. Virtual Network Embedding (VNE) addresses this problem by finding an effective mapping of the virtual nodes & links onto the substrate network.

For some scenarios, VNE has been studied in some detail in the network virtualization literature [1]. The problem was shown to be computationally intractable, but recent research has explored efficient heuristics to tackle the challenge.

**Motivation.** The VNE problem is traditionally formulated with the objective of maximizing network provider revenue by efficiently embedding incoming virtual network (VN) requests. This objective is subject to constraints, such as processing capacity on the nodes and bandwidth resource on the links.

A mostly unexplored perspective on this problem is providing some security assurances, a gap increasingly more acute. With the advent of network virtualization platforms [2], cloud operators now have the ability to extend their cloud computing offerings with virtual networks. To shift their workloads to the cloud, tenants trust their cloud providers to guarantee that their workloads are secure and available. Unfortunately, there is an increasing number of evidence that problems do occur, of both the malicious kind (e.g., caused by a corrupt cloud insider) or benign (e.g., a cloud outage) [3]. We thus argue that *security and dependability is becoming a critical factor that should be considered by virtual network embedding algorithms*.

To the best of our knowledge the only work that explores VNE security is the recent proposal by Liu et al. [4]. Despite its relevance, the authors fail to respond to the problems mentioned above: they do not contemplate dependability; and consider a single cloud provider, thus the model assumes complete trust in this entity.

**Contribution.** We propose a VN embedding solution that considers security and dependability as first class citizens. For this purpose, we introduce specific security constraints including, for instance, the possibility of a virtual machine attacking another virtual machine (e.g., a side-channel attack) or replay attacks on physical links. As substrate resources may fail, we also take into account dependability constraints, including the ability to tolerate failures, by ensuring that additional computing and communication resources are allocated during the process of embedding.

To further extend the resiliency properties of our solution, we assume a multiple cloud provider model (e.g., one based on nested virtualization [5]). We consider the coexistence of multiple clouds: both private, belonging to the tenant, and public, belonging to cloud providers. By not relying on a single cloud provider we avoid internet-scale single points of failures, avoiding cloud outages by replicating workloads across clouds. In addition, we can enhance security by leaving sensitive workloads in the tenant's private clouds.

**Solution.** We formulate this problem as a Mixed Integer Program (MIP). The *objective* is, as is common, to minimize the cost of embedding VN requests. We consider the typical flow conservation and resource capacity *constraints*. After defining specific security and dependability levels and demands, we further define the following additional constraints.

*Security*: a physical resource should guarantee at least the security level required by the virtual resource; physical resources should not host virtual resources that are potentially harmful to its operation; virtual resources that should not be co-hosted on the same physical resource as another virtual resource; a physical path should guarantee at least the security level required by the virtual link; sensitive virtual resources should not be hosted in public clouds.

*Dependability*: a physical resource should guarantee at least the replication level required by the virtual resource; the physical path should guarantee at least the replication level required by the virtual link.

Due to the complexity inherent to the embedding problem and the considerable size of the problem space in our MIP formulation, this solution is not efficient. We are currently investigating efficient greedy heuristics.

## References

[1] Fischer, A. et al. "Virtual Network Embedding: A Survey", IEEE Communications Surveys & Tutorials, 2013

[2] T. Koponen et al., "Network Virtualization in Multi-tenant Datacenters", NSDI'14

[3] Cloud Security Alliance, "The notorious nine cloud computing top threats in 2013"

[4] Liu et al. "Security-aware virtual network embedding", ICC'14

[5] M Ben-Yehuda, "The turtles project: design and implementation of nested virtualization", OSDI'10